\documentclass[aps,prd,preprint,amssymb,eqsecnum,nofootinbib,floatfix]{revtex4}
\usepackage{epsfig}
\usepackage{amsmath}
\usepackage{amsfonts}
\usepackage{bm}
\usepackage{revsymb}
\usepackage{graphicx,epsfig}
\usepackage{placeins}
\newcommand{\be}{\begin{equation}}
\newcommand{\ee}{\end{equation}}
\newcommand{\bea}{\begin{eqnarray}}
\newcommand{\eea}{\end{eqnarray}}
\newcommand{\bes}{\begin{subequations}}
\newcommand{\ees}{\end{subequations}}

\begin{document}
\interfootnotelinepenalty=10000

\title{Cosmological backreaction and spatially averaged spatial curvature}


\author{Eran Rosenthal, \'Eanna \'E. Flanagan}
\affiliation{Center for Radiophysics and Space Research, Cornell  University, Ithaca, New York, 14853}
\date{printed \today{} }

\begin{abstract}
It has been suggested that the accelerated expansion of the Universe is due to backreaction of small scale 
density perturbations on the large scale spacetime geometry.
While evidence against this suggestion has accumulated, it has not yet been definitively ruled out. Many investigations 
of this issue have focused on the Buchert formalism, which computes spatial averages of quantities in synchronous comoving
gauge. We argue that, for the deceleration parameter of this formalism to agree with observations, the spatial average of the 
three dimensional Ricci scalar (spatial curvature) must be large today, 
with an $\Omega_k$  in the range of $1 \le  \Omega_k  \le 1.3$.
We argue that this constraint is difficult to reconcile with observations of the location of the first Doppler peak of the CMBR.
We illustrate the argument with a simple toy model for the effect of backreaction, which we show is generically 
incompatible with observations.
\end{abstract}

\maketitle
\section{Introduction and summary}\label{intro}

Measurements of luminosity distance  as function of redshift  for 
type Ia supernovae, as well as measurements of inhomogeneities in the cosmic microwave background 
radiation, indicate that the expansion of the Universe is accelerating today \cite{Riess,Perlmutter,Bennett}. 
Explanations of this phenomenon usually involve either an introduction of "dark energy" -- 
a form of matter with negative pressure, or  a modification of general relativity.
Recently, a different explanation has been put forward \cite{Rasanen,Notari,Kolb1,Kolb2,lisch1,lisch2,Wiltshire,lnw,buchert2}, 
where the  acceleration is a consequence of subhorizon density perturbations.
According  to this idea, small scale cosmological density perturbations
evolve in a nonlinear manner to produce backreaction that affects the large scale 
spacetime geometry and modifies  the expansion of the Universe. 
This explanation is controversial and many authors  have argued that backreaction can not
explain the current acceleration of the Universe  \cite{sf,iw,bbr}.
Our viewpoint is that backreaction is likely to be too small  
to produce a significant modification to the large scale expansion of the universe. 
However, it deserves  to be investigated in detail since the backreaction 
explanation has not yet been definitively refuted.

To quantify the rate of expansion of an inhomogeneous  Universe, 
  Buchert \cite{buchert1,buchert2} introduced a particular method of taking a spatial average of 
the Einstein equations. He specialized to comoving synchronous gauge, and  
on each surface of constant time, denoted by $t$, he considers a spatial domain  $D(t)$ such that 
the boundary of $D(t)$ is comoving [i.e. $D(t)$ is independent of time in the synchronous comoving coordinates].
Defining  $V_D(t)$ to be the proper volume of this domain, the effective scale factor $a_D(t)$
is given by 
\[
\frac{4\pi}{3} a^3_D(t)=V_D(t)\,.
\]
By averaging the Einstein equations for an irrotational dust Universe, 
Buchert derived the following evolution equations for $a_D(t)$ 
\begin{eqnarray}\label{ad}\label{aprime}
&&\left(\frac{a'_D}{a_D}\right)^2 =\frac{8\pi}{3}\rho_{eff} \,,\\ \label{adbprime} 
&&- \frac{a''_D}{a_D} = \frac{4\pi}{3}(\rho_{eff}+ 3p_{eff})\,,
\end{eqnarray}
where prime denotes differentiation with respect to cosmic time $t$,
 and throughout this paper we use geometrized units where  $G=c=1$. Equations (\ref{aprime},\ref{adbprime}) 
have the same form as the Friedmann equations, except that their sources are an effective 
density and an effective pressure, $\rho_{eff}$ and  $p_{eff}$, respectively, which are defined by
\begin{eqnarray}\label{rhoeff}
\rho_{eff}\equiv\langle \rho \rangle_D -\frac{1}{16\pi}(\langle R_3 \rangle_D + \langle Q \rangle_D) \,,\\ \label{peff}
p_{eff} \equiv -\frac{1}{16\pi}(\langle Q \rangle_D -\frac{1}{3} \langle R_3 \rangle_D)\,.
\end{eqnarray}
Here  $\rho $ denotes the matter density, and $R_3$ denotes the spatial three-dimensional Ricci scalar. 
 The brackets $\langle...\rangle_D$ denote an average over the domain $D(t)$, for example
\[
\langle R_3 \rangle_D\equiv \int_D R_3 \sqrt{ \det(g_{ij})} dV\ /\ \int_D \sqrt{\det(g_{ij})}dV=
V_D^{-1}\int_D R_3 \sqrt{ \det(g_{ij})} dV\,,
\]
where ${\det(g_{ij})}$ denotes the determinant of  the spatial 3-dimensional induced 
metric, and $dV$ denotes the  three-dimensional coordinate volume-element. 
The quantity denoted $\langle Q \rangle_D$ is defined by
\begin{equation}
\langle Q \rangle_D \equiv \frac{2}{3}\langle(\theta-\langle \theta \rangle_D)^2 \rangle_D-\langle \sigma_{\alpha\beta} \sigma^{\alpha\beta} \rangle_D\,.
\end{equation}
Here $\theta$ denotes the dust expansion parameter and $\sigma_{\alpha\beta}$  denotes the shear tensor.
Notice that the quantity $\langle Q \rangle_D$  vanishes for a homogeneous and isotropic Universe, but 
becomes nonzero if one includes density perturbations. Furthermore,  
Eq. (\ref{adbprime}) implies that  a sufficiently large value of $\langle Q \rangle_D$ could produce 
a negative value for 
$\rho_{eff}+ 3p_{eff}=\langle \rho \rangle_D -({1}/{4\pi})\langle Q \rangle_D$, 
and by virtue of Eq. (\ref{adbprime}) 
give rise to an accelerated expansion  $a''_D>0$. 

While the effective scale factor $a_D$ is a mathematically well defined quantity, its relation  to  cosmological observations 
is not completely clear. The physical interpretation of $a_D$ faces three main difficulties. 
First, $a_D$  is a quantity defined on a  spacelike hypersurface, and so, in general, 
it  can not be directly related  to cosmological observations which are determined by quantities 
on the  past lightcone of the  observer. 
Second, the time evolution of $a_D$ does not provide sufficient information 
to allow calculation of cosmological observables 
such as luminosity distance as function of redshift, which requires a metric for its calculation.   
Third, defining a quantity related to a constant time hypersurface
is  somewhat  arbitrary, since one is free to choose a different  time coordinate that defines a different  spacetime 
foliation. Despite these difficulties,  it has been argued that if the Buchert formalism predicts 
an effective declaration parameter $q_D$ which is approximately $-1/2$, then it is likely 
that the predicted value of the actual declaration will also be large and negative. In this 
paper we will adopt this point of view, and ignore the above mentioned difficulties with the interpretation of $a_D$.

In Ref. \cite{lisch1}, Buchert's formalism is used to calculate the evolution of $a_D$ in a perturbed 
Friedmann Robertson Walker (FRW) irotational dust Universe. In particular the term $\langle Q \rangle_D$  which presumably 
drives the accelerated expansion of the Universe is calculated to 
second order in perturbation theory, and is found to be a boundary term, depending only on the metric perturbations 
on the boundary of the domain $D(t)$. This result generalizes other previous computation using
Newtonian cosmological perturbation theory, which calculates a quantity related to $\langle Q \rangle_D$  
which is also found to be a boundary term \cite{aei}.
However,  there is a difficulty in reconciling this property  of $\langle Q \rangle_D$ at second-order
with the interpretation of  $\langle Q \rangle_D$ as the source of backreaction.  
To see this, suppose that the  Universe is spatially compact, 
and that the domain $D$ is chosen to be the complete  space. 
In this case any boundary term must vanish identically, and can have no affect on the time evolution of 
the Universe. While there is no observational evidence for a compact Universe, the fact that  an FRW Universe has a particle 
horizon implies that a noncompact Universe 
is observationally indistinguishable from a spatially compact one as long as 
the scale of compactness  is larger then the observer's particle horizon at decoupling. This bound  translates into  
  a lower bound on  the scale of compactness today of about  twice the size of the horizon.
We are  therefore free to choose the scale of compactness to be 
roughly twice the horizon size today, without changing anything we can measure.
This implies that   $\langle Q \rangle_D=0$ for $D\approx 2\times{\rm horizon}$.
Now Buchert's formalism is valid for all choices of $D$ and gives no guidelines as to what choice of 
$D$ to make. This ambiguity is part of the overall problem of relating the Buchert formalism 
to observations. Yet, it seem plausible  that the correct answer (if it exist) should be roughly 
$D\approx {\rm horizon\  size\  today}$. 
It seems reasonable that the value  of  
$\langle Q \rangle_D$ should  not change much if we reduce $D$ from $2\times{\rm horizon}$ 
to roughly the horizon size, and if so,  
for this choice of $D$, the Buchert formalism predicts that 
expansion of the Universe is unaffected by the vanishing $\langle Q \rangle_D$ at second-order.
Nevertheless, there remains the possibility that third order and higher order perturbations could 
produce a large backreaction effects \cite{Notari} which can not be represented 
by a boundary term so the backreaction issue is not settled.

In this paper we argue that general considerations suggest  
that it is hard to reconcile a large cosmological  backreaction 
 described by the Buchert formalism with   
observational constraints coming from measurements
 of luminosity distance and angular-diameter distance as functions of redshift. 
Observations of the the first Doppler peak of the CMBR together with baryon acoustic 
oscillation and supernovae data has been used to severely constrain the 
spatial curvature of a  ${\rm\Lambda CDM}$   Universe. By combining these observations 
with the assumptions that the dynamics of Universe is  
governed by the FRW metric and that the effect of density perturbations is negligible 
it has been found that spatial curvature satisfies  
$\Omega_K=-0.0052\pm0.0064$ ($68\%$ CL) \cite{WMAP}, where  $\Omega_K=-R_3(t_0)/(6H_0^2)$, 
$t_0$ denotes the current time and $H_0$ denotes the current Hubble rate.
In this paper we shall consider a more general theoretical framework 
that include perturbations and possibly large backreaction.  
One might expect that these observations should also place constraints on  
the average curvature $\langle R_3 (t_0)\rangle_D$.
In order for the Buchert formalism to reproduce the desired 
accelerated expansion from backreaction alone, 
it must  have a significant averaged curvature with 
 $0.975 \le  \Omega_k  \le 1.294 $ (see Sec. \ref{k0}), where
we have defined  $\Omega_k\equiv-\langle R_3 (t_0)\rangle_D/[6H_D^2(t_0)]$,  and denoted
the effective Hubble rate by  $H_D=a'_D/a_D$. This large averaged curvature seems to be  hard to reconcile 
with the flat Universe implied by observations. 

One possible avenue for evading this observational constraint in 
spatial curvature, suggested in Ref. \cite{buchert2}, is the fact that in the Buchert formalism
the effective energy density in the spatial curvature need not have the standard scaling $\propto a^{-2}$.
It is not clear whether the strong constraints coming from CMBR are more sensitive to the low redshift
curvature or high redshift curvature. If the constraint principally applies to high redshift curvature, 
then an evolving curvature that is negligible at high redshift could evade the CMBR constraints. 
In this paper  we shall argue that this avenue for evading the constraint is unlikely. 
Any  nonstandard time evolution of  the spatial curvature is quite constrained, since at  high 
redshifts  the density perturbations evolve linearly and the 
 Universe is accurately described by  a weakly perturbed CDM Universe. 
Non-standard time evolution must therefore be confined to 
the low redshifts, where nonlinear effects are presumably  important. 
However, in this regime the spatial  curvature 
is constrained by the requirement that the backreaction formalism 
 reproduces the correct  luminosity distance as function of redshift that agrees with supernovae data. 
These observations constrain the time evolution of the 
metric. Therefore, a nonstandard time evolution of the curvature in this regime would require 
 a nonstandard evolution of the metric such  that  supernovae data observations are reproduced despite  the
large spatial curvature. 
In this scenario  the full time evolution of the metric has two regimes. 
In the first low redshift regime, the metric evolves 
in a highly non-standard manner, and in the second high redshift regime,  
it evolves according to a standard weakly perturbed CDM Universe. 
The difficulty  that it is not guaranteed  
that this time evolution reproduces  the correct angular  power spectrum as measured by WMAP. 

In this paper we construct a simple toy model that illustrates the observational 
difficulties that arise in models with a large value of averaged spatial curvature today, even 
allowing for nonstandard evolution of that curvature. 
For this purpose, we adopt the point of view of the Buchert backreaction formalism, and assume that 
  we can replace the actual spacetime geometry by a set of averaged quantities.  
To be able to make predictions, we assemble these quantities and construct an 
 averaged metric that allows us to calculate observables.
Here we should make the following remarks.

First, the spatial 
averaged curvature in the Buchert formalism need not be dominated 
by low spatial frequency components, it may be mostly high 
spatial frequency components. Nevertheless, CMBR photons traveling along our
past lightcone experience some sort of average curvature along  their way. 
While this average is different from that of the Buchert formalism, it is plausible that  they are not too much different. 
We will not address this issue in this paper. Second, in this paper we shall calculate an average spatial curvature
using an expression for an averaged metric. However,  this calculation is in general different
 from an average of the curvature of the true metric. We shall ignore this discrepancy in this paper.

We construct  the following averaged metric toy model 
\begin{equation}\label{metric}
d{s}^2=a^2(\eta)\left[-d\eta^2+\frac{dr^2}{1-k(\eta)r^2}+r^2d\Omega^2\right]\,.
\end{equation}
The  time coordinate $t$ is related to $\eta$ by $dt^2=a^2 d\eta^2$. 
It should be emphasized that this metric should be thought of as an averaged metric and so it does not have 
to satisfy the Einstein field equations. 
Here, $a(\eta)$ and $k(\eta)$ are certain functions of the conformal time $\eta$, 
where we set the present value of the scale factor to unity $a(\eta_0)=1$. 
This averaged metric is designed to allow for  a time evolution of the averaged spatial
curvature to mimic what is presumably produced by backreaction.
Notice that for every constant time hypersurface the 
induced three dimensional metric obtained from 
(\ref{metric}) coincides with a corresponding induced 
three-metric of an FRW constant time slice, and so 
this three-metric is isotropic and homogeneous about every point. 
However, for a generic function $k(\eta)$, 
the overall four-dimensional spacetime is not maximally symmetric. 
Finally, we should mention here that 
after completing this work we  learned that the form (\ref{metric}) 
 of an averaged metric has been 
suggested before in Refs. \cite{buchert2}, 
see also Refs. \cite{mwmk,Rasanen2}.

Recently the toy model (\ref{metric}) was studied in detail 
by Larena et. al. \cite{buchert3}.  This study claims 
that there is a good agreement between 
this toy model and data from WMAP and supernovae observations, while we 
reach the opposite conclusion. 
We believe that the reason for this 
discrepancy originates from the fact that 
 Larena et. al.  use a different expression for  the 
redshift in terms of the scale factor and the function $k(\eta)$.  
In their study  it is argued that under some approximation 
the relation between redshift and scale factor 
is the standard $1+{z}\propto a^{-1}$  relation [see  their Eq. (31)]. 
Using the standard definition of redshift (\ref{rshift1}) we show  
that the nonstandard time evolution of the spatial curvature 
significantly changes this relation, and the correct relation
is  given by Eq. (\ref{rshift3}). As we show, this nonstandard expression 
for the redshift 
has a significant effect on the calculation of observables in this model.

Our goal in this paper is to confront the model (\ref{metric}) 
with observations. For this purpose we calculate the luminosity distance $D_L(z)$ and 
angular diameter distance $D_A(z)$ as functions of redshift, 
using the metric (\ref{metric})
 and compare the results with observational constraints\footnote{In practice, 
it is sufficient to calculate $D_A(z)$, since $D_L(z)$ can be obtained 
from  the relation $D_L(z)=(1+z)^2 D_A(z)$ which is valid in any spacetime, see Refs. \cite{Ellis,Etherington}. } . 
We start by choosing a set of functions $k(\eta)$ parametrized by two 
parameters [see Eq. (\ref{ketadef}) below]. 
For each set of values of the parameters we then choose $a(\eta)$ 
 to enforce the equation $D_L(z)=D^{{\rm\Lambda}CDM}_L(z)$ at low redshifts,  
where $D^{{\rm\Lambda}CDM}_L(z)$ is the luminosity distance derived from a $\Lambda CDM$ 
FRW model with parameter values agreeing with supernovae observations. 
This equation is enforced up to a maximum redshift. 
Using this requirement we calculate $a(\eta)$  for this low redshift part of the spacetime.
We focus attention only on those metrics in which the 
the spatial curvature vanishes at a large redshift. 
Once the averaged spatial curvature vanishes  the backreaction effect should vanish  as well, 
 and so in this high redshift regime we assume that $a(\eta)$ follows  the standard evolution of a $CDM$  
cosmology without a cosmological constant. 
Using this law of evolution we calculate the function $a(\eta)$ for the remaining part of spacetime. 
Once we have  calculated $a(\eta)$, 
we use the metric (\ref{metric}) to  compare  the 
 characteristic angular scale of the CMBR power spectrum as derived from 
our model with  observation of  WMAP.  
We find that generically the characteristic angular scale of our model 
is at odds with WMAP  observations. 

This paper is organized as follows. 
In Sec. \ref{findfunctions} we explain in detail how we determine the 
the functions $a(\eta)$ and $k(\eta)$ of our model . 
In Sec. \ref{firstpeak} we calculate the sound horizon that determines the location 
of first CMBR peak in our model. 
In Sec. \ref{results}  we  explore various values 
of the parameters which determines 
the function $k(\eta)$ and describe the results.

\section {Construction of the backreaction model} \label{findfunctions}

To be able to interpolate between an initially 
 vanishing averaged spatial curvature that  corresponds to a weakly perturbed 
FRW Universe, and  a current large  
value of averaged spatial curvature needed for the backreaction picture,
we  assume that the function  $k(\eta)$ in the metric (\ref{metric}) 
takes the following form 
\begin{equation}\label{ketadef}
k(\eta) = \begin{cases}
 H_0^2 \bar{k}\frac{f^2} {f^2+1} & , \eta\ge\bar{\eta} 
 \\ 0 & , \eta\le\bar{\eta}  \end{cases}
\end{equation}
where
\[
f\equiv\frac{H_0(\eta - \bar{\eta})} {w}    \,.
\]
Here $H_0=(a^{-1} \frac{da}{dt})_{t_0}$ is the value of the Hubble constant today, and 
$\bar{k}$, $w$ and $H_0 \bar{\eta}$ are dimensionless parameters. 
The parameter $\bar{\eta}$ marks the conformal time of the transition between a conformally flat spacetime and 
a conformally curved spacetime, and $w$ governs the rapidity of this transition.
Using the definition $\Omega_k\equiv-\langle R_3 (t_0)\rangle_D/[6H_D^2(t_0)]$ 
together with  Eq. (\ref{ketadef}) and identifying $H_0$ with $H_D(t_0)$ we find that  
\begin{equation}\label{omegak}
\Omega_k=-\bar{k}\left(1+\frac{w^2}{(\eta_0 - \bar{\eta})^2}\right)^{-1}\,,
\end{equation}
where $\eta_0$ is the value of the conformal time today. 
Below we explore various values for the parameters $w$ and $\bar{\eta}$, 
while $\bar{k}$ is determined from the requirement that the Buchert formalism gives 
an equation of state parameter of dark energy near $-1$ [see  Sec. \ref{k0}].

To calculate  the scale factor $a(\eta)$, we demand that 
the luminosity distance as function of redshift,  $D_L^{model}(z)$, in our model matches observational data.   
Since the luminosity distance of a ${\rm\Lambda}CDM$  cosmology  matches observational data 
we impose    
\begin{equation}\label{fitdl}
D^{model}_L(z)=D^{{\rm\Lambda}CDM}_L(z)\,,
\end{equation} 
where  throughout the superscripts 'model' and $'{\rm\Lambda}CDM'$ refer to our model and to a flat ${\rm\Lambda}CDM$ cosmology, respectively.
We impose the condition (\ref{fitdl}) only at low redshifts, in the range of values of $\eta$ given by $\eta\ge \bar{\eta}$.
At high redshifts, we switch to imposing the Friedmann equation, since backreaction should be negligible at
high redshifts.
The evolution of $a(\eta)$ for $\eta\le \bar{\eta}$ is determined from 
an Einstein- de Sitter model for which the scale factor satisfies
\[
\dot{a}=h \sqrt{ a}\,,
\]
 where an overdot denotes differentiation with respect to $\eta$. We determine the constant $h$ 
by demanding continuity of $\dot{a}$ at the transition time $\bar{\eta}$. 
The solution of this equation is given by
\begin{equation}
a(\eta)=\left[\frac{h}{2}(\eta-\bar{\eta})-\sqrt{a(\bar{\eta})}\right]^{2}\ ,\ \eta\le \bar{\eta} \,.
\end{equation}
where  we used the continuity of   $a({\eta})$  at $\eta=\bar{\eta}$ .

\subsection{Matching luminosity distances as function of redshift}

In this section we describe how we  calculate $a(\eta)$  in practice from the matching  requirement (\ref{fitdl}) .  
In a general spacetime the luminosity distance $D_L(z)$ is related to the angular diameter distance 
$D_A(z)$ by \cite{Etherington,Ellis} 
\begin{equation}\label{dlda}
D_L(z)=D_A(z)(1+z)^2\,.
\end{equation}
Using  Eq. (\ref{dlda}), the matching  requirement (\ref{fitdl}) takes the form of 
\begin{equation}\label{fitda}
D^{model}_A(z)=D^{{\rm\Lambda}CDM}_A(z)\,.
\end{equation} 
In  a flat ${\rm\Lambda}CDM$ cosmology the right hand side of Eq. (\ref{fitda}) is given by
\begin{equation}\label{dafrw}
D^{{\rm\Lambda}CDM}_A(z)=\frac{1}{(1+z)H_0}\int_0^z [\Omega_m(1+z')^3+\Omega_\Lambda]^{-1/2}dz'\,.
\end{equation}
Here the parameters $\Omega_m$ and $\Omega_\Lambda$ satisfy $\Omega_m+\Omega_\Lambda=1$, and the 
 contribution from radiation energy-density has been neglected since we  confine 
the discussion to the epoch after recombination. 

We now consider the left hand side of Eq. (\ref{fitda}) and derive an expression for  $D^{model}_A$.
Suppose that an observer views a sizeable distant object (e.g. a distant galaxy or a structure of the CMB anisotropy) 
that has  a  transverse proper cross sectional area $\delta A$, and 
subtends a small solid angle $\delta{\Omega}$. From these quantities the observer can determine 
the angular diameter distance 
\[
D_A=\sqrt{\frac{\delta A}{\delta\Omega}}\,.
\]
Since the wavelength of the electromagnetic radiation  is typically much  smaller than  the spacetime curvature, we 
 can safely use the geometric optics approximation and describe 
the electromagnetic radiation as a bundle of light rays that trace a congruence of null geodesics. 
We assume that  the light rays converge at  an event $p$ at the location of the observer,  
which is chosen to be at the origin, so  that $r(p)=0$ and  $\eta(p)=\eta_0$. 
Since the metric  (\ref{metric}) is isotropic about the origin, the light rays trace 
radial null geodesics from the source at $r(\eta)$, where $\eta<\eta_0$, to the observer;   
 and  the angular diameter distance is given by 
\begin{equation}\label{daar}
{D}^{model}_A=a(\eta)r(\eta) \,.
\end{equation}
Substituting Eqs. (\ref{dafrw}) and (\ref{daar}) into Eq. (\ref{fitda}) and differentiating with respect to $\eta$ gives 
\begin{equation}\label{fit2}
H_0 \frac{d}{d\eta}[(1+z)a r]=\frac{dz}{d\eta} [\Omega_m(1+z)^3+\Omega_\Lambda]^{-1/2} \,.
\end{equation}
Our goal to solve Eq. (\ref{fit2}) for $a(\eta)$. As a preliminary step, we first calculate 
 $r(\eta)$ and $z(\eta)$ and then substitute these functions into Eq. (\ref{fit2}). 

The  calculation of  $r(\eta)$  for radial null geodesics follows directly from the metric (\ref{metric}), which  
gives $\dot{r}^2=1-k(\eta)r^2$, where dot denotes differentiation with respect to $\eta$. 
Later we shall assume that $k(\eta)<0$ and so $\dot{r}^2>0$ implies that the null geodesics have  no turning points. 
Since $\eta$ is a monotonically decreasing function of $r$  we have 
\begin{equation}\label{rdot}
\dot{r}=-\sqrt{1-k(\eta)r^2} \,.
\end{equation}
Below we use Eq. (\ref{ketadef}) to specify $k(\eta)$ and solve Eq. (\ref{rdot}) numerically 
together with the initial condition $r(\eta_0)=0$. 

We now consider the calculation of the redshift $z(\eta)$.
By definition the redshift is given by
\begin{equation}\label{rshift1}
1+{z}=\frac{{({k}^\alpha {u}^\beta {g}_{\alpha\beta})}_{source}}{{({k}^\alpha {u}^\beta {g}_{\alpha\beta})}_{observer}}\,,
\end{equation}
where $k^\alpha$ is the 4-momentum of the photon, and $u^\alpha$ is the 4-velocity of the 
cosmological fluid.
For simplicity we assume that both the observer and the source have  
four-velocities of the form ${u}^\alpha=a^{-1}\delta^\alpha_\eta $ meaning that their  
peculiar velocities  vanish.
We normalize $k^\alpha$ by demanding that ${k}^\alpha {u}^\beta g_{\alpha\beta}=-1$ at the observer. 
Some simplification is gained by  considering a conformal transformation  of the form 
\begin{equation}\label{conformal}
g_{\alpha\beta}=a^2 \hat{g}_{\alpha\beta} \ \ , \ \ {k}^\alpha=a^{-2}\hat{k}^\alpha\,.
\end{equation}
Combining our choices for normalization and velocities with 
 Eqs. (\ref{rshift1}) and (\ref{conformal}) gives 
\begin{equation}\label{rshift2}
{z}=a^{-1}\hat{k}^\eta-1\,.
\end{equation} 
The conformal null  vector field $\hat{k}^\eta$  satisfies a geodesic equation in the conformal spacetime where the metric is $\hat{g}_{\mu\nu}$,
and so it is independent of $a(\eta)$. 
Using this geodesic equation together with Eq.(\ref{rdot})  we obtain 
\begin{equation}\label{ketadot}
(\hat{k}^\eta)^{-1}  \frac{d}{d\eta}\hat{k}^\eta =-\frac{r^2\dot{k}}{2(1-kr^2)}\,.
\end{equation}
Integrating Eq. (\ref{ketadot}) and using  Eq. (\ref{rshift2}) we find that the red shift is given by 
\begin{equation}\label{rshift3}
{z}=a^{-1}e^{1/2\int_{\eta}^{\eta_0} r^2\dot{k} {(1-kr^2)}^{-1} d\eta' }-1 \,.
\end{equation}
Notice that for $k={\rm const}$ Eq. (\ref{rshift3}) reduces to the FRW relation $z+1\propto 1/a$. 
Eq. (\ref{rshift3}) is at odds with Eq. (30) of Ref. \cite{buchert3} which seems to be inconsistent with 
the standard definition of redshift (\ref{rshift1}). 
Below we solve for  $r(\eta)$ by specifying $k(\eta)$ and solving Eq. (\ref{rdot}). Using $r(\eta)$ we evaluate the integral in Eq. (\ref{rshift3}) 
and obtain  $z(\eta)$. Both $r(\eta)$ and $z(\eta)$  are then substituted into  Eq. (\ref{fit2}) which is solved to give $a(\eta)$. Finally 
$a(\eta)$ and $r(\eta)$ are inserted into Eq. (\ref{daar}) to obtain ${D}^{model}_A(\eta)$, and this is combined with $z(\eta)$ to obtain 
the angular diameter distance $D_A^{model}(z)$ as function of redshift $z$. 
All these calculations are done numerically.

\subsection{Constraining the toy model parameters}\label{k0}

In this section we use observational constraints  on the cosmological parameters 
to  place constraints on the parameters of our toy model. 
The calculation is based on Buchert's formalism  \cite{buchert1}  which was summarized in Sec. \ref{intro}.
In this formalism  the equation of state parameter of dark energy is given by 
\begin{equation}\label{eqofstate}
w_{de}(\eta)=\frac{p_{de}(\eta)}{\rho_{de}(\eta)}\,,
\end{equation}
where   $\rho_{de}$ denotes the dark energy density, and  $p_{de}$ denotes 
the dark energy pressure. These quantities are related to the effective 
density  $\rho_{eff}$ and effective pressure $p_{eff}$, which are given by  Eqs. (\ref{rhoeff}) and (\ref{peff}), trough 
the relations  $\rho_{eff}=\langle \rho \rangle_D+\rho_{de}$ and $p_{eff}=p_{de}$. 
Eq. (\ref{eqofstate}) together with Eqs. (\ref{rhoeff},\ref{peff}) give 
\begin{equation}\label{wde}
w_{de}=\frac{-(1/3)\langle R_3\rangle_D+\langle Q\rangle_D}{\langle Q\rangle_D+\langle R_3\rangle_D}\,.
\end{equation}
The current value of this parameter is in the range  $-1.1 \le w_{de}(\eta_0) \le -0.9$ \cite{woodetal}.  Using  this constraint together with 
Eq. (\ref{wde}) gives
\begin{equation}\label{qr}
-\frac{57}{17}\langle Q  (\eta_0)\rangle_{D}\le \langle R_3 (\eta_0) \rangle_{D} \le -\frac{63}{23}\langle Q  (\eta_0)\rangle_{D} \,.
\end{equation}
We define the effective deceleration parameter  by 
\begin{equation}\label{qdef}
q_D=-\frac{a''_D}{a_D H_D^2}\,,
\end{equation}
and substitute Eq. (\ref{adbprime}) into Eq. (\ref{qdef}) and use  Eqs. (\ref{rhoeff}) and (\ref{peff}). This gives
\begin{equation}\label{qd}
q_D(\eta_0)=\frac{1}{2}\Omega_{m(D)}-\frac{\langle Q(\eta_0)\rangle_D}{3H_D^2}\,,
\end{equation}
where $\Omega_{m(D)}=8\pi\langle \rho \rangle_D/3H_D^2$. 
We demand that this expression be equal to the 
current deceleration  of a standard flat ${\rm \Lambda}CDM$ cosmology, where the deceleration parameter is given by 
\begin{equation}\label{q0frw}
q_{\rm \Lambda CDM}(\eta_0)=\frac{1}{2}\Omega_m-\Omega_\Lambda\,.
\end{equation}
where  $\Omega_m$ and $\Omega_\Lambda$ are the ${\rm \Lambda CDM}$ densities of matter and dark energy, respectively.
Assuming that we can substitute   $\Omega_m$ in place of $\Omega_{m(D)}$ 
 we find from Eq. (\ref{qd}) and Eq.(\ref{q0frw}) that 
\begin{equation}\label{qom}
\frac{\langle Q(\eta_0)\rangle_D}{3H_D^2}=\Omega_{\Lambda}\,,
\end{equation}
The 5-year WMAP data \cite{WMAP} reveals that  the dark energy density  is 
in the range  $\Omega_{\Lambda}=0.742\pm 0.030$. We use the WMAP constraint on  $\Omega_{\Lambda}$
and  Eqs. (\ref{qom}), (\ref{qr}) together with the definition  $\Omega_k\equiv-\langle R_3 (t_0)\rangle_D/[6H_D^2(t_0)]$, 
to  obtain
\begin{equation}\label{findk0}
0.975 \le  \Omega_k  \le 1.294 \,.
\end{equation}
By combining $ \Omega_k\approx1.1$ together with  Eq.  (\ref{omegak}) and Eq. (\ref{findk0}) we 
determine  the parameter $\bar{k}$ once the parameters 
$w$ and $\eta_0$ have been specified.

\section{The characteristic angular scale of the  CMBR power spectrum}\label{firstpeak}

The position of the peaks of the WMAP  angular power spectrum is set by the characteristic 
angular scale $\theta^{WMAP}$ defined by
\begin{equation}\label{thetadef}
\theta^{WMAP}\equiv\frac{D_H(z^*)}{D_A(z^*)}\,,
\end{equation}
where  $D_H(z^*)$ denotes the sound horizon at the redshift of decoupling, where throughout 
 we use the notation $*$ to refer to  the redshift of decoupling. 
For a flat $\Lambda CDM$ cosmology the observed WMAP power spectrum 
is  consistent with the angular scale (\ref{thetadef}) at a percent level of accuracy 
(Assuming that the sound horizon is consistent 
with a $\Lambda CDM$ cosmology 
the  5-year WMAP data \cite{WMAP} yields a  comoving  angular diameter distance given by  
$d_A^{obs}(z^*)=14115^{+188}_{-191}{\rm Mpc}$). 
To see if our toy model is able to reproduce the 
characteristic   angular scale which is consistent with observations  we calculate the ratio 
\be\label{angelratio}
\chi=\frac{\theta^{model}}{\theta^{WMAP}}=
\frac{ D_H^{model}(z^*) D_A^{\Lambda CDM}(z^*)}{D_A^{model}(z^*)D_H^{\Lambda CDM}(z^*)}\,.
\ee
To calculate $\chi$ we need to calculate the ratios 
$D_A^{\Lambda CDM}(z^*)/D_A^{model}(z^*)$
and   $D_H^{model}(z^*) / D_H^{\Lambda CDM}(z^*)$.
The calculation of the ratio of angular diameter distances 
follows  from the method described in Sec.  \ref{findfunctions}.  
We now discuss calculating the ratio of the sound horizons at decoupling, for this 
we follow Ref.  \cite{Weinberg}.  Early on our model coincides with a cold dark matter cosmology. 
Therefore,  prior to decoupling the sound speed in the plasma of baryons and photons  
is given by $v_s=[3(1+R)]^{-1/2}$, where $R\equiv 3\rho_B/4\rho_\gamma$; and  
$\rho_B$ and $\rho_\gamma$ denote the baryon density and the photon density, respectively.
At this epoch  these densities are assumed to be approximately homogeneous, and 
the sound horizon takes the form of 
\begin{equation}\label{soundh}
D_H^{model}=a^*\int_0^{t^*}\frac{dt}{a\sqrt{3(1+R)}}\,.
\end{equation}
To evaluate this integral we need to know the time evolution of the relevant densities  
before  decoupling. 
In  our model the  baryons are  assumed to be comoving,  so that  
 the evolution of the baryon density (and the matter density) at the location of the
observer traces the evolution of a three dimensional volume element at $r=0$,   
which gives 
\begin{equation}\label{rhob}
\rho_B=\rho_{B0} a^{-3}\ ,\ t<t^*\,.
\end{equation}
Here and throughout the subscript $0$ denotes 
a quantity evaluated at the observer today, for example  $\rho_{B0} \equiv\rho_B(\eta_0,r=0)$.  
The evolution of the photon density (and the radiation density) is  that of a black-body and  so
it  is  determined by the temperature. Therefore,  irrespective of any spacetime symmetry we have 
\begin{equation}\label{rhogamma}
\rho_\gamma=\rho_{\gamma 0} (1+z)^4\  ,\ t<t^*\,.
\end{equation}
Recall that in our model there is a nonstandard relation between the redshift and the scale factor.  
For this reason it is instructive to introduce the quantity $\alpha\equiv a^{-1}(1+z)$.  
Prior to the growth in curvature,  i.e. at a conformal time  $\eta$ where  $\eta<\bar{\eta}$,   
Eq. (\ref{rshift3}) implies  that $\alpha$ is a constant, given by 
\begin{equation}\label{alpha}
\alpha =e^{1/2\int_{\bar{\eta}}^{\eta_0} r^2\dot{k} {(1-kr^2)}^{-1} d\eta' }\,.
\end{equation}
For  a $\Lambda CDM$ Universe we have  $\dot{k}=0$ and so we recover $\alpha=1$. 
Using Eqs. (\ref{rhob},\ref{rhogamma},\ref{alpha}) we obtain 
\be\label{rr0}
R=R_0 \alpha^{-4}a\,.
\ee
Using this equation together with Eq. (\ref{soundh}) we can write the sound horizon as 
\be\label{soundh2}
D_H^{model}=R^*\int_0^{t^*}\frac{dt}{R\sqrt{3(1+R)}}\,.
\ee
The integration is carried out by noting that $dt=dR/HR$, 
$H=\sqrt{(8\pi/3)(\rho_M+\rho_R)}$, where $\rho_M$ and $\rho_R$ denote the 
matter density and the radiation density, respectively. 
Introducing the notation $R^*_{\Lambda CDM}=R_0 (1+z^*)^{-1}$ and $R_{EQ}=3\rho_{R0} \rho_{B0} /4\rho_{M0} \rho_{\gamma_0}$  
we obtain 
\be\label{soundhratio}
\frac{D_H^{model}(z^*)}{D_H^{\Lambda CDM}(z^*)}=\alpha^3 
\frac{f(\alpha,R^*_{\Lambda CDM},R_{EQ})}{f(1,R^*_{\Lambda CDM},R_{EQ})}\,,
\ee
 where
\[
f(\alpha,R^*_{\Lambda CDM},R_{EQ})=
\ln\left(\frac{\sqrt{1+R^*_{\Lambda CDM}\alpha^{-3}}+\sqrt{R_{EQ}+R^*_{\Lambda CDM}\alpha^{-3}} }{1+\sqrt{R_{EQ}}}\right)\,.
\]
Here we assumed  that  the observer in our  toy our model Universe would measure the same densities today as 
an observer placed in a $\Lambda CDM$ Universe. For a standard $\Lambda CDM$ Universe \cite{Weinberg} we have 
$R^*_{\Lambda CDM}=0.62$, $R_{EQ}=0.21$. 
Eq. (\ref{soundhratio}) implies that  at decoupling 
the sound horizon calculated form  our toy model is  different 
from the sound horizon in a $\Lambda CDM$ Universe.  
This suggests  that for generic parameters  $(\bar{\eta},w)$ 
our model would give rise to  a shift in the locations of the   peaks of the TT 
power spectrum determined by the ratio ${\theta^{model}}/{\theta^{WMAP}}$ given by 
Eq. (\ref{angelratio}). An exception to this could arise only in rare occasions where the ratio 
$D_H^{model}(z^*) / D_H^{\Lambda CDM}(z^*)$ in Eq. (\ref{angelratio})
is exactly compensated by the ratio  
$ D_A^{\Lambda CDM}(z^*) / D_A^{model}(z^*)$ in this equation.  

\section{Results}\label{results}

\begin{figure}
\begin{center}
\includegraphics[width=.45\textwidth]{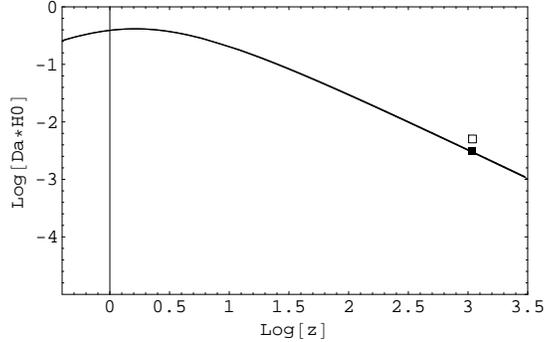}
\caption{Log-log plot of the angular diameter distance as 
function of redshift,  for a ${\rm \Lambda CDM}$ Universe and the backreaction 
toy  model with parameters $(H_0\bar{\eta},w)=(-1.6,1.2)$ which corresponds to a transition 
redshift of $z(\bar{\eta})=1.92$. 
For these parameters the two graphs of   $D_A^{model}(z)$ and   $D_{\Lambda CDM}^{model}(z)$   coincide.  
The filled box  indicates the WMAP measured  angular diameter distance at decoupling  $D_A^{obs}(z^*)$ 
assuming that the sound horizon is consistent with a ${\rm \Lambda CDM}$ Universe. 
The empty box   indicates this distance assuming that the sound horizon is consistent with 
the toy model. }
\label{fig1}
\end{center}
\end{figure}

We explored the two dimensional parameter space  $(\bar{\eta},w)$ of the metric function $k(\eta)$  and used 
the method described in Sec. \ref{findfunctions}  to  determine the functions $k(\eta)$ and $a(\eta)$. We then used  
Eq. (\ref{daar}) and Eq. (\ref{rshift3})  to determine the angular diameter distance at decoupling 
 $D_A^{model}(z^*)$, where  $z^*=1090.51\pm0.95$  \cite{WMAP}. 
To calculate  the ratio  $\chi={\theta^{model}}/{\theta^{WMAP}}$ 
we substituted  the value of $D_A^{model}(z^*)$ and the value of  $D^{{\rm\Lambda}CDM}_A(z^*)$ [given by  Eq. (\ref{dafrw})]
into  Eq. (\ref{angelratio})  and used Eq. (\ref{soundhratio}). 
We have found that for generic values of  $(\bar{\eta},w)$ the estimator $\chi$ is different from unity by more than a percent  
indicating a discrepancy between the backreaction toy-model and observations. An example of this discrepancy is 
demonstrated in Fig. \ref{fig1} showing 
the angular diameter distance as function of redshift   for a ${\rm \Lambda CDM}$ Universe and the backreaction 
toy  model for  parameters  $(H_0\bar{\eta},w)=(-1.6,1.2)$. 
For these parameters the  
graphs of  $D_A^{model}(z)$ and   $D_{\Lambda CDM}^{model}(z)$  coincide.  However, the WMAP value of the 
angular diameter distance at decoupling is  
model dependent and is  
in agreement only with a ${\rm \Lambda CDM}$ Universe.  
There is a small region in the parameter space where  $(H_0\bar{\eta},w)\approx(-0.8,0.8)$
and  $z(\bar{\eta})\approx 1$  where the ratio 
$D_H^{model}(z^*) / D_H^{\Lambda CDM}(z^*)$ in Eq. (\ref{angelratio})
is exactly compensated by the ratio  
$ D_A^{\Lambda CDM}(z^*) / D_A^{model}(z^*)$ in this equation.
While this part of the parameter space of our toy model is not ruled out,  
we are not aware of any observation that 
support a sudden growth of the averaged curvature  at $z\approx 1$.  

\begin{figure}
\begin{center}
\includegraphics[width=.45\textwidth]{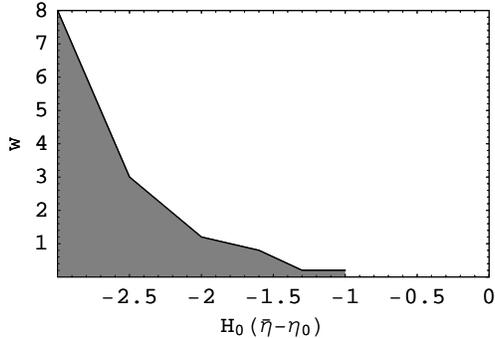}
\caption{The toy-model parameter space $(H_0\bar{\eta},w)$, showing the 
domain (gray) where the redshift becomes a  non-monotonic function of the conformal time coordinate $\eta$.}
\label{fig2}
\end{center}
\end{figure}

\begin{figure}
\begin{center}
\includegraphics[width=.45\textwidth]{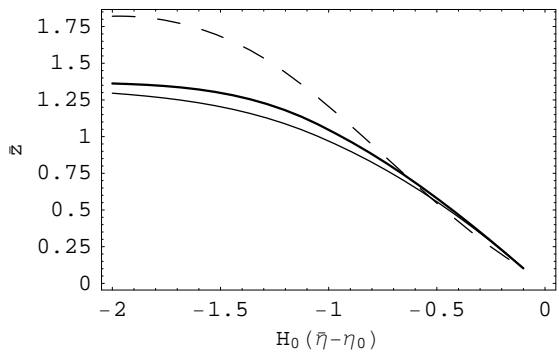}
\caption{The  transition redshift $\bar{z}$  
as function of the model parameter $\bar{\eta}$ for different values of the parameter $w$.
Showing $w=0.05$ (thin line), $w=0.2$ (thick line), $w=0.8$ (dashed line).}
\label{fig3}
\end{center}
\end{figure}

Another difficulty  that showed up only in  a portion of the toy-model parameter-space is  that 
 the redshift can become  a non-monotonic function of the conformal time coordinate 
 $\eta$. The portion of the parameter space which suffers from this difficulty is shown in Fig. \ref{fig2}.
Finally, Fig \ref{fig3} shows the transition redshift $\bar{z}\equiv z(\bar{\eta})$ as function of the parameters
of the toy-model. For model parameters where the redshift is a monotonic function of $\eta$ 
an increase in $w$ (for a fixed $\bar{\eta}$) normally implies a decrease in $\bar{z}$. Notice, however, that most of the graphs 
are in the portion parameter space where the redshift is not a monotonic function of $\eta$.

\section{Conclusions}\label{conclusions}

In this paper we studied a toy model of a backreaction mechanism.  
In this model the averaged spatial curvature grows at low redshifts 
so that the expansion of the Universe presumably induced by backreaction  
could be consistent with supernovae data. 
In the high redshift regime,  we assumed that Universe  
 evolves according to a standard weakly perturbed CDM Universe. 
We showed that this model  alters the predictions for the 
sound horizon at decoupling and that it is generically inconsistent with the 
 power spectrum as measured by WMAP.

\acknowledgments 
This work was supported by NSF  grants, PHY 0457200 and PHY 0757735.

\newcommand{\apjl}{Astrophys. J. Lett.}
\newcommand{\aap}{Astron. and Astrophys.}
\newcommand{\cmp}{Commun. Math. Phys.}
\newcommand{\grg}{Gen. Rel. Grav.}
\newcommand{\lr}{Living Reviews in Relativity}
\newcommand{\mnras}{Mon. Not. Roy. Astr. Soc.}
\newcommand{\pr}{Phys. Rev.}
\newcommand{\prsl}{Proc. R. Soc. Lond. A}
\newcommand{\ptrsl}{Phil. Trans. Roy. Soc. London}

\end{document}